\documentclass[apj]{emulateapj}
\usepackage{comment}
\usepackage{enumerate}
\usepackage[T1]{fontenc}
\usepackage{pslatex}
\usepackage{epstopdf}

\slugcomment{The Astrophysical Journal, 687:L000-L000, 2008 November 10}

\def\deg{^\circ }

\begin{document}


\title{A Spectacular H$\alpha$ Complex in Virgo:
Evidence for a Collision Between M86 and NGC 4438
and Implications for the Collisional ISM Heating of Ellipticals}

%
%
\author {Jeffrey D. P. Kenney\altaffilmark{1},
Tomer Tal\altaffilmark{1},
Hugh H. Crowl\altaffilmark{2},
John Feldmeier\altaffilmark{3},
George H. Jacoby\altaffilmark{4}}
%
\altaffiltext{1}{\scriptsize 
Yale University Astronomy Department,
P.O. Box 208101, New Haven, CT 06520-8101 USA jeff.kenney@yale.edu, tomer.tal@yale.edu }

\altaffiltext{2}{\scriptsize 
Dept. of Astronomy, LGRT-B 619E, University of Massachusetts, 710 North Pleasant St., 
Amherst, MA 01003-9305, hugh@astro.umass.edu}

\altaffiltext{3}{\scriptsize
Department of Physics \& Astronomy,
Youngstown State University, One University Plaza, Youngstown, Ohio, 44555, jjfeldmeier@ysu.edu}

\altaffiltext{4}{\scriptsize
WIYN  Observatory, 950 N. Cherry Avenue, Tucson, Arizona 85719, jacoby@noao.edu}


\shorttitle{H$\alpha$ Evidence for M86 Collision}
\shortauthors{KENNEY ET AL.}

\begin{abstract}
Deep wide-field H$\alpha$+[NII] imaging around the Virgo cluster giant
elliptical
galaxy M86 reveals a highly complex and disturbed ISM/ICM.
The most striking feature is a set of H$\alpha$ filaments which clearly connect
M86 with the nearby disturbed spiral NGC~4438 (23$'$=120 kpc projected away), providing
strong evidence for a previously unrecognized collision between them.
Spectroscopy of selected regions show a fairly smooth velocity gradient between M86 and
NGC 4438, consistent with the collision scenario.
Such a collision would impart significant energy into the ISM of M86,
probably heating the gas and acting to prevent the gas from cooling to form
stars. We propose that cool gas stripped from NGC 4438 during the collision and
deposited in its wake is heated by shocks, ram pressure drag, or thermal conduction,
producing most of the H$\alpha$ filaments.
Some H$\alpha$ filaments are associated with the well-known ridge of
bright X-ray emission to the NW of the nucleus, suggesting that the collision is
responsible for peculiarities of M86 previously ascribed to other effects.
M86 is radio-quiet, thus AGN heating is unlikely to play a significant role.
The M86 system has implications for understanding the role of gravitational
interactions in the heating of the ISM in ellipticals,
and how collisions in clusters transform galaxies.
\end{abstract}


\section {Introduction}

The warm ionized plasma of elliptical galaxies is an important phase of the ISM
that traces evolutionary processes and galactic interactions.
H$\alpha$ emission from this ISM phase has been observed in many
ellipticals \citep{trinch91,mac96},
and has sometimes been interpreted to be the result of
``cooling flows'', in which the hot X-ray emitting plasma
radiates, cools down, and sinks toward the galactic center \citep{fab94}.
However since
the mass in colder gas and newly formed stars in most galaxies is far less
than simple cooling models predict
\citep{pet03}, a heating source is required to limit the net cooling rate.
Heating by radio-loud AGN has received much attention recently
and may be an important process \citep{croton06,bru07,best07,bow06}
for limiting cooling flows and the continued stellar growth of massive galaxies.
However another heating source which may be important,
especially in radio-quiet galaxies, is gravitational interactions
\citep{spa04,db08}.
Correlations between H$\alpha$ emission and
merger tracers such as dust and tidal tails provide
suggestive evidence for a connection between
gravitational interactions and the warm ionized gas in some ellipiticals
\citep{spa04},
but clear evidence for such connections have been limited.
In this paper we present compelling new evidence for
a gravitational interaction origin for the H$\alpha$ emission
around the nearby elliptical M86.

The giant elliptical M86, near the core of the Virgo cluster,
is the most massive galaxy in a large
group or sub-cluster now merging with the rest of the Virgo cluster
\citep{sch99}.
It is blueshifted by 1350 km/s with respect to M87 at the heart of
the cluster, and thought to be falling into the Virgo cluster from behind.
We adopt a distance to M86 of 17.5 Mpc \citep{jac90, ms98, mei07},
which is $\sim$1 Mpc further than the distance to M87.
M86 has an unusually complex ISM.
There is a bright "plume"  of X-ray emission from hot gas,
beginning 3$'$= 15 kpc projected north of the nucleus and extending at least 10$'$= 50 kpc  projected 
to the NW \citep{for79}.
Several authors have suggested that this is
a tail of gas stripped from M86 as it falls into the Virgo ICM
\citep{for79, whi91, ran95, ran08},
although other X-ray properties are not consistent with
a simple stripping picture \citep{fin04}.
M86 also has unusual clouds of HI \citep{bre90, li01}
and far-infrared \citep{sti03}  emission,
whose origin has not been understood.
Based on  X-ray and FIR properties,
\citet{fin04} and \citet{sti03}
suggested galaxy collisions may be involved,
but could not identify clear evidence for any collision.
M86 is a very weak radio source \citep{dc78, hum80, fab89}
thus the ISM complexities are unlikely to be related to AGN activity,
but are instead likely due to interactions in a group and cluster
environment.

\begin{figure*}
\plotone{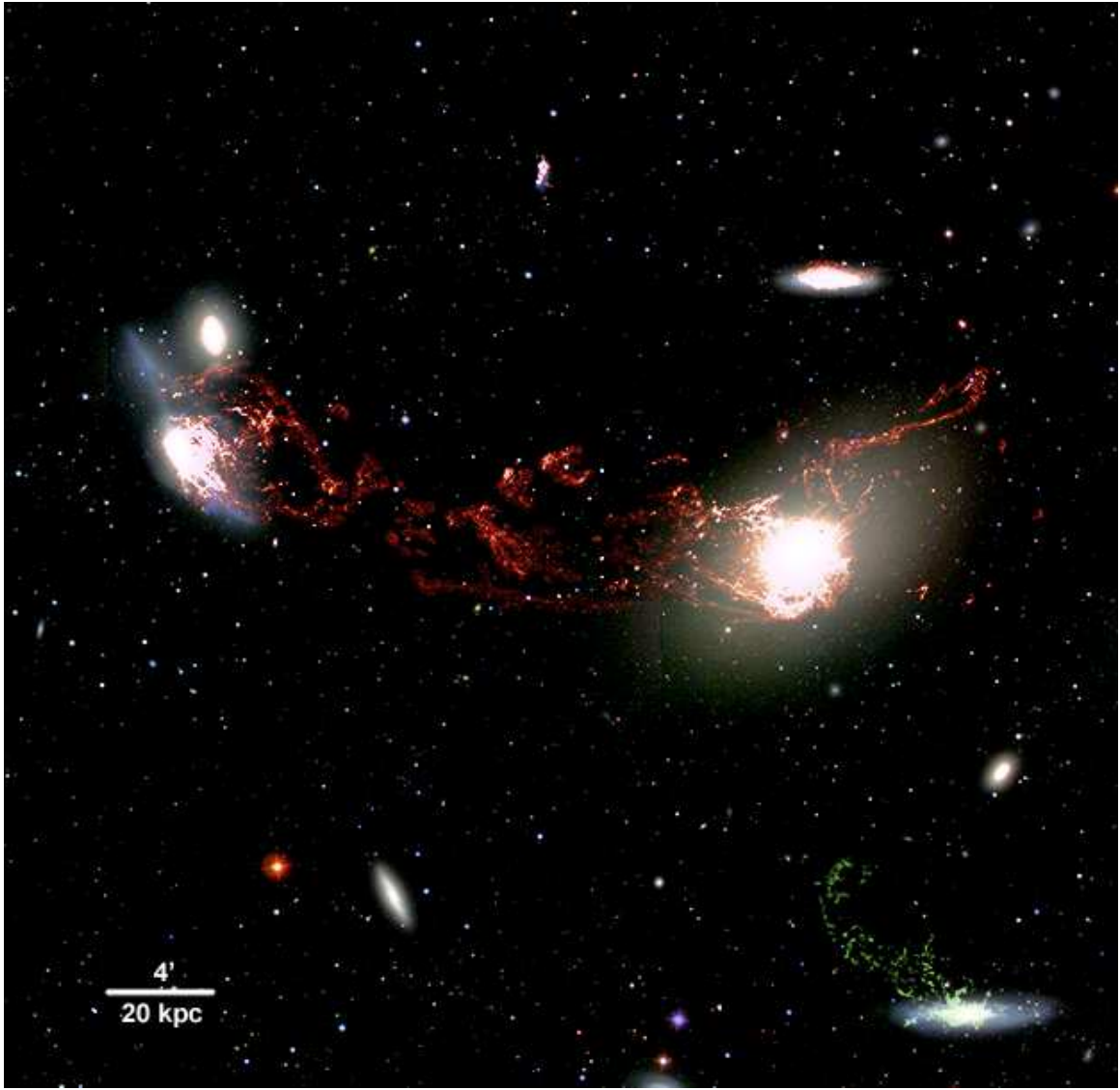}
\caption{H$\alpha$+[NII] image of M86 region superposed on a
color SDSS gri image. The H$\alpha$ image is stretched to highlight the faint emission.
The ``low-velocity'' (<500 km/s) H$\alpha$+[NII] emission is colored red,
and the ``high-velocity'' (>2000 km/s) H$\alpha$+[NII] emission is colored
green. The low-velocity emission is attributed to gas stripped from NGC~4438 
in a collision with M86 and subsequently heated.
High-velocity emission is observed near NGC~4388 in the lower right,
although it does not extend all the way to M86, so it is unclear whether it is related.
}
\end{figure*}

In this paper we present a deep wide-field H$\alpha$+[NII] image
around M86 which shows a highly complex and disturbed ISM/ICM,
including striking new evidence for a collision between M86 and the
nearby heavily disturbed spiral NGC~4438.
Both galaxies have been the subject of numerous individual studies, but this is
the first evidence that has clearly linked them together.
We propose that most of the peculiarities of the ISM in M86 and NGC 4438 are due
to this high-velocity collision,
and discuss the implications for understanding the role of gravitational
interactions in the energetics and evolution of the ISM in ellipticals.
\\
\section{Observations and Data Reduction}

The M86 region was observed with the MOSAIC imager on the KPNO 4m Mayall 
telescope, in order to search for warm gas which could reveal clues about interactions.
Exposures were taken using three
filters: the Harris R (k1004, $\lambda$/$\delta$$\lambda$ = 6514\AA /1511\AA) broad-band filter and two H$\alpha$
narrow-band filters - rest frame (k1009, 6575/81) and redshifted (k1010, 6521/80).
We imaged the system with 3 different pointing centers, and with 2 different H$\alpha$
filters,
in order to have spatial coverage out to the nearby galaxies NGC 4438, NGC 4388,
and M84, and wavelength  coverage corresponding to both low (-1280 to 2380 km/s
FWHM) high (780 to 4430 km/s) velocities. Total exposure times per
field/velocity were 2000 sec for the R-band, and 7000-8000 sec for H$\alpha$.
Images were processed and mosaiced together in a standard way, except for
additional flat fielding corrections using masked image frames,
and a pupil correction for the H$\alpha$ images.
The H$\alpha$ image superposed on an SDSS gri image \citep{adel07} is shown in Figure 1,
the R-band image is shown in Figure 2a, and
the H$\alpha$ image is shown in Figures 2b and 2c.

We obtained spectroscopy at 24 selected positions near M86
with the SparsePak integral field unit (IFU) on the WIYN telescope.
SparsePak \citep{ber04} is a 90-fiber array  which
loosely covers an 80$''\times$80$''$ field of view, with 5$"$ diameter fibers.
The fibers feed the Bench Spectrograph
with a 860 line/mm grating, providing a
spectral resolution of 1.6\AA = 80 km/s and coverage from 5923-6862 \AA .
Each of the positions was observed for 1-2 hrs.
While the emission lines are strong and clearly detected in regions close to
the center of M86, the lines are weak in some of the outer positions, and for
these
we co-add the spectra from several adjacent fibers to measure the emission
lines.
We detect both the H$\alpha$ and the 6583 [NII] lines in these outer clouds,
but cite the velocity of the H$\alpha$ line, since it is usually stronger.
Representative velocities from the spectroscopy are indicated on Figure 2c.
Full imaging and spectroscopy results will be presented in a later paper.
\\
\section {Results \label{results}}

The H$\alpha$ morphology of M86 is very complex, as shown in Figure 2b.
There are filamentary features with a range of morphologies, but very few
small compact HII region-like knots (except near NGC~4438),
and no obvious FUV sources \citep{gil07},
therefore very little star formation.
We identify four major regions where the morphological
characteristics of the H$\alpha$ features appear to be distinct.

1. {\bf The inner 1-2$'$ of M86.}
The strongest H$\alpha$ emission originates in the central 1-2$'$ of M86,
where it forms an asymmetric complex extending 2$'$ to the S of the nucleus,
but only ~1' to the N.
The unusual  H$\alpha$ morphology in this region was previously
shown by \citet{trinch91}.
Some of the central H$\alpha$ emission  is located in the
same region
as an unusual off-nuclear cloud of HI emission \citep{li01},
which extends from the nucleus to 2$'$ S (see Figure 2b).
The southern edge of the HI cloud is spatially coincident with
semi-circular arc of H$\alpha$ which defines the southern limit of 
the ``eastern'' H$\alpha$ filaments (described in 2.)
suggesting they are physically associated.
This region shows a
disturbed velocity field, with all velocities within 250 km/s of
the central velocity of -240 km/s.

2. {\bf Near NGC 4438.}
Unusual H$\alpha$ filaments were previously detected in NGC~4438,
extending westward as far as 2$'$ from the nucleus \citep{kw93,ken95}.
However the new, deeper observations show H$\alpha$ emission
with a much larger extent to the west, including
a loop of emission extending ~5$'$ from NGC 4438,
apparently connecting to the filaments located between the 2 galaxies.
Only a small fraction of the total H$\alpha$ luminosity
of NGC~4438 is in the form of small discrete knots
consistent with HII regions \citep{ken95}. A GALEX UV study also
shows that the total star formation rate in this disturbed spiral
is modest \citep{bos05}.

3. {\bf Loop feature NW of M86.}
Extending 2-8$'$ (10-40 kpc  projected) from M86 to the NW at PA=315$\deg$
is a remarkably well-defined figure 8-shaped H$\alpha$ loop.
It has a roughly radial alignment,
with total extent of 6$'$ in the radial direction,
but only 1-2$'$ in the tangential direction.
This feature includes a linear filament 5$'$=22 kpc projected  in length,
and only 0.1$'$ in width implying an aspect ratio of 50:1,
which suggests a minimum of turbulence in this part of M86's ISM.
This H$\alpha$ loop is spatially coincident with the bright NW X-ray plume
(Figure 2b) and an associated HI cloud, suggesting that they are all related.
Unlike the eastern filaments,
this feature is blueshifted with respect to M86,
with velocities of -230 to -450 km s$^{-1}$.

4. {\bf Features extending eastward from M86 toward NGC 4438.}
There are a remarkable series of features east of the M86 nucleus,
which clearly extend to the large disturbed spiral galaxy NGC 4438
located 23$'$=120 kpc  projected away.
These features reach to within 1-2$'$ of the M86 nucleus, and are
centered about a well-defined axis with a PA=82$\pm$1 deg
which passes through the centers of both galaxies.
There are several distinct features between the 2 galaxies, including
a roughly triangular feature ~1-5$'$ east of the M86 nucleus,
and a curved arc of emission roughly halfway between the galaxies.
There is a large range in velocities in the triangle feature close to 
(1-2$'$ E of) M86, but otherwise
there is a fairly smooth velocity gradient between M86 and NGC 4438,
with -240 km s$^{-1}$ for M86,
-150 km s$^{-1}$ for the arc between the 2 galaxies, 
-40 km s$^{-1}$ for a cloud between the arc and NGC4438,
and +70 km s$^{-1}$ for NGC4438.
The continuity of the  kinematics and morphology
imply a physical assocation between M86 and NGC~4438.

The total H$\alpha$+[NII] luminosity of the M86-NGC4438 complex
is  1.3$\times$10$^{41}$ erg/s. Of this total,
32\% arises within 2$'$ of M86, 50\%  within 5$'$ of N4438,
12\% between the 2 galaxies, and 5\% from the NW figure 8 loop.
The H$\alpha$+[NII] luminosity of M86 is in the top 20\% 
of nearby bright ellipticals and S0's \citep{mac96}, and
is the 2nd largest among Virgo ellipticals
with a luminosity $\sim$2.5 times lower than M87 in the center of the cluster.

\section{Discussion}

\begin{figure*}
\plotone{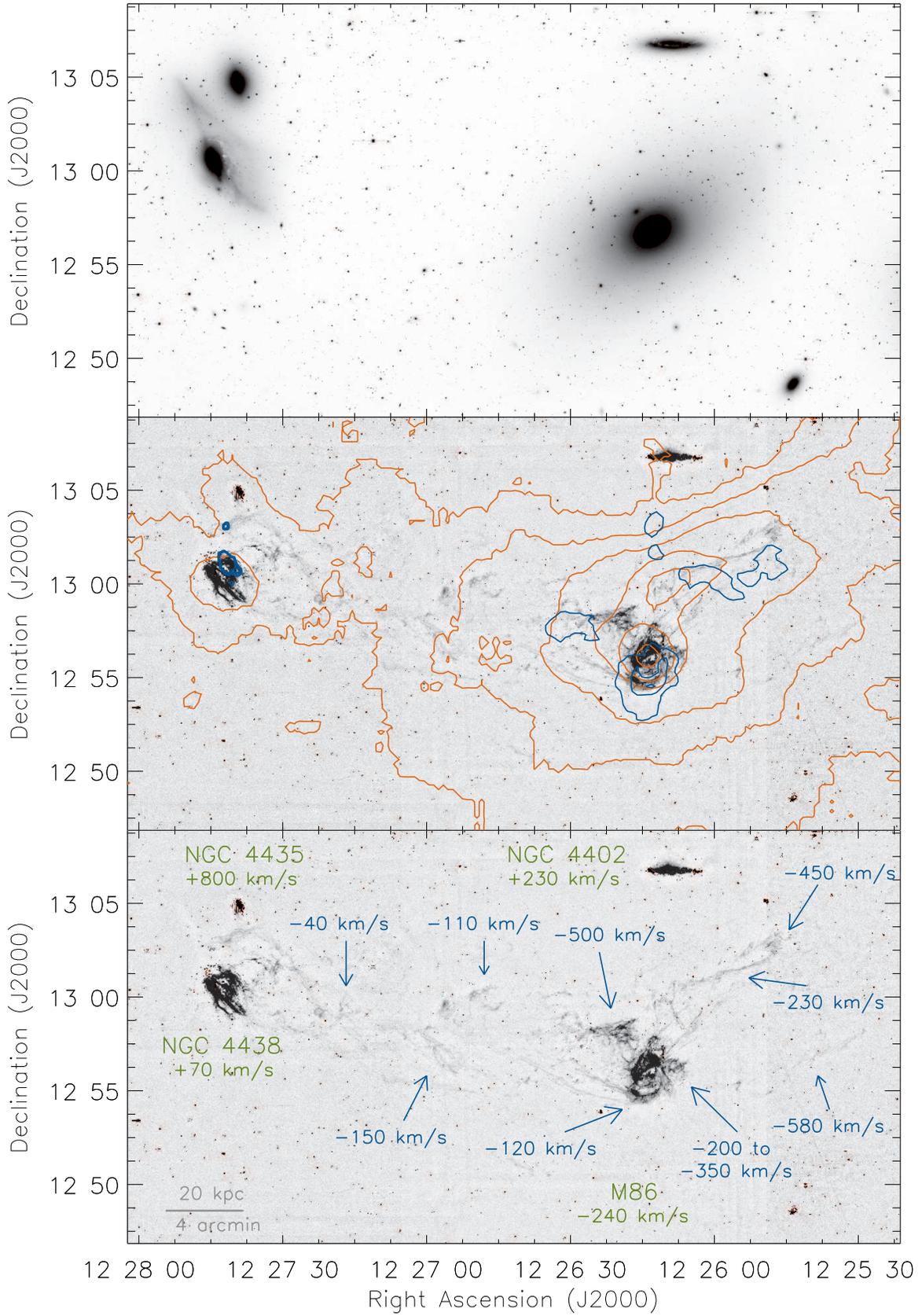}
\caption{a.) R-band image of M86 and NGC 4438 region,
shown with logarithmic stretch.
b.) H$\alpha$+[NII] image of M86 and NGC 4438 region,
shown with a logarithmic stretch to highlight the fainter features.
ROSAT X-ray contours (red) and VLA HI contours (blue) are  superposed.
H$\alpha$, HI and X-ray peaks are closely associated in several regions.
NGC~4438 has both  X-ray and HI emission offset toward the west, associated
with some of the H$\alpha$ filaments.
c.) H$\alpha$+[NII] image of M86 and NGC 4438 region, with H$\alpha$
velocities of selected features identified.}
\end{figure*}

We propose that most of the H$\alpha$ features mark a gaseous debris trail and
wake caused by the passage of NGC 4438 through the ISM of M86.
This would make the M86-NGC~4438 system the
nearest example of a recent high-velocity collision between a large elliptical
and large spiral.

NGC~4438 (Arp120) is the most heavily disturbed large spiral in the
Virgo cluster. The outer stellar disk is disturbed, forming extraplanar tidal
arms, indicating a strong gravitational interaction.
The ISM of NGC 4438 is even more disturbed than the stars. In particular, all
ISM tracers in NGC 4438 are displaced to the west
of the stellar disk \citep{ken95}, on the side toward M86, where they appear
to connect to the large scale H$\alpha$ filaments linking NGC~4438 to M86.
NGC~4438 is very HI-deficient, with an HI mass of 4$\times$10$^8$ M$_\odot$,
only $\sim$5-15\% of the expected HI mass for a spiral galaxy of its size
\citep{gks83,ken95,sol02,gav05}.
This suggests that 5$\pm$2$\times$10$^9$ M$_\odot$
of gas has recently been lost from
NGC~4438, or converted into other forms. While some ram pressure stripping
may also be expected from NGC~4438's passage through the general Virgo ICM
\citep{vol05}, the results in this paper (further developed below)
suggest that most of the ISM of
NGC 4438 was stripped from its stellar disk by the ram pressure caused by
its passage through M86's ISM. Previous papers
\citep{com88,ken95,vol05}
have suggested that the enormous disturbances
to NGC 4438 are due to a collision with the SB0 galaxy NGC 4435, which lies
very nearby (only 5$'$ away) on the sky. However NGC 4435 shows no clear signs
of disturbance itself. Its stellar distribution and its small
nuclear gas and dust disk appear normal. With a radial velocity 700 km/s higher
than that of NGC 4438, NGC 4435 is probably unrelated to the disturbances of
NGC 4438.
The H$\alpha$ imaging and kinematics presented in this paper, showing a clear
connection
between
M86 and
NGC 4438, provides strong evidence that most of the disturbances to NGC 4438
were caused by a collision with M86.
For an average plane-of-sky velocity of 800 km s$^{-1}$ (less than peak velocity)
the time since closest approach in the collision would be 100 Myr,
a timescale consistent with the stellar populations in the outer disk of 
NGC~4438 \citep{bos05}.

Cool gas stripped from NGC~4438 is a reasonable explanation for the
unusual clouds of HI near M86 \citep{li01}.
Figure 2b shows that the two of the largest HI clouds in M86 lie within the
eastern H$\alpha$ filament system that
connects the 2 galaxies, suggesting that they originated from NGC 4438, and
were stripped from it during the collision. Indeed, the lower edge of the
southern HI cloud is spatially coincident with the lower H$\alpha$ filaments
which connect M86 and NGC~4438.
A 3rd HI cloud is associated with the NW X-ray plume \citep{li01}.
The total HI mass in M86 of 2$\times$10$^8$ M$_\odot$ \citep{bre90},
is a small fraction of the HI missing from NGC~4438,
and the remainder may have been heated and
entered another phase within the halo of M86.
The mass of H$\alpha$-emitting gas in uncertain, as it depends on the
poorly known gas densities, but small, probably less than 10$^7$ M$_\odot$
\citep{mac96}.

While it is pretty clear that the eastern filaments connecting M86 and NGC~4438
are caused by the collision, it is less clear whether the
NW  H$\alpha$ loop and associated X-ray ridge are due to the collision.
Previous authors have suggested that this prominent high-surface brightness
X-ray ridge in the NW is
a tail of gas stripped from M86 as it falls into the Virgo ICM
\citep{for79, whi91, ran95, ran08}.
However, the X-ray morphology of M86 is unlike those in other Virgo
and Fornax E galaxies with evidence for ram pressure stripping, which show
sharp, flattened leading surface brightness edges consistent with
bowshocks and centered trailing tails \citep{mac05a,mac06}.
There are a number of observations in M86 which are
not consistent with a simple stripping picture,
and more easily understood in a collision scenario:
the presence of an extended X-ray halo which appears mostly
relaxed, unlike the complexities in the core \citep{fin04},
and the presence of discrete clouds of HI and FIR emission
at various locations within the halo, including within the NW ridge.
The NW gas ridge may be a ram pressure stripped gas tail disturbed by the collision.
Alternately, we propose that the NW X-ray ridge, H$\alpha$ loop and HI cloud are
related to the incoming part of NGC~4438's trajectory through the M86 halo,
during which 
most of NGC~4438's cool gas was stripped and subsequently heated.
The blueshift of this gas relative to M86, which is opposite that
the eastern filaments, is consistent with the following collision scenario:
NGC 4438 approaches M86 from behind and as it accelerates toward M86,
it is blueshifted. 
Gas in a ram pressure stripped tail should be accelerated toward the cluster mean
velocity, and therefore redshifted, opposite from what is observed.
\footnote{
There is also the possibility of another collision, between M86 and a 3rd
galaxy, the spiral NGC~4388 
which lies 10$'$ south of M86.
A long tail of HI gas stripped from NGC~4388 extends up towards M86
(Oosterloo \& van Gorkom 2005).
However, the line-of-sight velocity of NGC~4388 and its HI tail range from
2100-2500 km/s,
much higher than M86, NGC~4438, and the H$\alpha$ emitting gas.
We have 
detected the high-velocity extended emission (shown in green in Fig 1)
previously detected within ~5$'$ of NGC~4388  \citep{yos04},
but no high velocity H$\alpha$ emission coincident with the northern
parts of the HI tail close to M86. This suggests that the HI tail of gas
stripped from NGC 4388
may be unrelated to M86, although this possibility warrants further investigation.}

Cool gas stripped from NGC 4438 and deposited in its wake will be moving within the 
hot gaseous halo of M86, and subject to heating.
The source of excitation of the warm ionized gas
is unlikely to be UV photoionization from stars,
since there are no FUV sources detected from GALEX \citep{gil07}
and the H$\alpha$ image shows very few compact HII regions.
The spatial correlation of HI, H$\alpha$, and X-ray emission in M86 (Figure 2b)
is suggestive of warm ionized gas at the interface between cold and hot gas,
and suggests that the excitation of the H$\alpha$ filaments is due to
shocks, collisional excitation from ram pressure drag \citep{db08},
or thermal conduction from the hot halo \citep{spa04,con01}.


The kinetic energy of the gas stripped from NGC 4438 in the collision is likely
a significant heating source for the ISM of M86.
The ratio of X-ray to optical luminosity in M86 is significantly higher than the average value 
for ellipticals \citep{ds07}, perhaps a consequence of heating in the collision.
In order for NGC 4438 to pass close to the center of M86, and escape as far
away as 120 kpc  projected with a relative velocity of more than 300 km s$^{-1}$, it must have
been moving much faster, exceeding the escape speed,
which is $\sim$1000 km s$^{-1}$ at R=10 kpc, when it passed through the center.
Assuming 5$\times$10$^9$ M$_\odot$ of gas stripped from NGC 4438 during its
passage through M86's gaseous halo,
and a relative velocity at closest approach of 1000 km s$^{-1}$,
the kinetic energy of the stripped ISM is 5$\times$10$^{58}$ erg,
a significant fraction of which will be absorbed by the ISM of M86 in the
collision. This is similar to the thermal energy nkTV, of
M86's ISM, $\sim$10$^{59}$ erg,
where we have assumed kT=1.0 keV, an average density of n=2$\times$10$^{-3}$ cm$^{-3}$
over the volume V of a sphere with radius of 15$'$=73 kpc projected, roughly the area of
the ROSAT X-ray emission \citep{ran95, ran08}.
Thus the collision will be a significant energy source for
M86's ISM, and might act to prevent the gas from cooling to form stars.
While there has been significant attention paid recently to the heating effects
of AGN in early type galaxies,  \citep{croton06,bru07}
gravitational heating through collisions, mergers and gas accretion
can be a significant heating source \citep{db08}.
While M86 is certainly an extreme case, 
the effects are easier to study in extreme cases, thus 
this system will have implications for understanding the
ISM and ICM heating which occurs in the hierarchical assembly of galaxies and
clusters.



\acknowledgements

We are grateful to the staffs of the KPNO and WIYN observatories for
their assistance obtaining the data,
Jacqueline van Gorkom for kindly rereducing and providing HI maps, and
Chris Mihos, Patrick Durrell, Riccardo Giovanelli, Pieter van Dokkum, 
and Jacqueline van Gorkom for helpful discussions.

\facility
{\it Facilities: } \rm \facility{Mayall }, \facility{WIYN }

\end {document}